\documentclass{svjour3}                     
\smartqed  
\usepackage{fix-cm}
\usepackage{graphicx}
\usepackage{amsmath}
\usepackage{amssymb}
\usepackage{overpic}
\usepackage{inputenc}
\journalname{Few-Body systems}
\begin{document}

\title{Study on triple-hadron bound states with Gaussian expansion method} \thanks{Grants or other notes
}


\author{Tian-Wei Wu         \and
        Li-Sheng Geng 
}


\institute{Tian-Wei Wu \at
              School of Physics, Beihang University, Beijing 102206, China \\
           \and
           Li-Sheng Geng \at
         School of Physics \& Beijing Key Laboratory of Advanced Nuclear Materials and Physics, \\
          Beihang  University, Beijing 102206, China\\
School of Physics and Microelectronics,\\
Zhengzhou University, Zhengzhou, Henan 450001, China\\
          \email{lisheng.geng@buaa.edu.cn}
}

\date{Received: date / Accepted: date}
\maketitle

\begin{abstract}

In recent years, more and more exotic hadronic states have been discovered successively. Many of them can be explained as hadronic molecules, such as  $D_{s0}^*(2317)$, $X(3872)$, and $P_c$ pentaquark states. Analogous to the formation of nuclei, we study three-body hadronic molecules with the Gaussian expansion method, and predict the existence of the $DDK$, $\Xi_{cc}\Xi_{cc}\bar{K}$, and $BB\bar{K}$ bound states, which are likely to be found in the current and updated facilities.
\end{abstract}

\section{Introduction}
\label{intro}
In recent years,  many new hadronic states beyond the traditional quark model have been found, which are collectively referred to as exotic hadronic states.
Among them, quite a large amount can be interpreted as hadronic molecules, that is, bound or resonant states formed by two hadrons via residual strong interactions, such as the  $D_{s0}^*(2317)$ and $P_c$ pentaquark states, which can be well interpreted as $DK$ and $\Sigma_c\bar{D}^{(*)}$ molecular states~\cite{Geng:2010vw,Liu:2019tjn}.

This picture of two-body molecular states  can be extended to three-body hadronic systems by an accurate few-body  method, the Gaussian expansion method~\cite{Hiyama:2003cu}. In this work, we study the $DDK$ system based on the $DK$ interaction and one boson exchange model for charmed mesons, and predict the existence of a $DDK$ molecular state~\cite{MartinezTorres:2018zbl,Wu:2019vsy,Huang:2019qmw,Pang:2020pkl}. Utilizing heavy quark symmetry~\cite{Neubert:1993mb}, the $B\bar{K}$ and $\Xi_{cc}\bar{K}$ interactions can be related to the
$DK$ interaction, from which the $BB\bar{K}$ and $\Xi_{cc}\Xi_{cc}\bar{K}$ systems are also studied.

These few-body hadronic molecular states can decay in a specific way~\cite{Huang:2019qmw,Wu:2020job}, which is expected to be observed in the current or upgraded experimental facilities. If such new hadronic states composed of multiple hadrons are found  experimentally, the picture of two-body hadron molecules can be verified.

\section{Two-body interactions}
\label{Interactions}

In order to solve these 3-body systems, namely the $DDK$, $BB\bar{K}$, and $\Xi_{cc}\Xi_{cc}\bar{K}$ systems, we have to first specify the two-body interactions. 
All of these three systems are composed of  two identical hadrons (mesons or baryons) and a kaon (antikaon). In the following, we use $AAB$ to represent these three systems, with $A$
the $D$, $B$ mesons or $\Xi_{cc}$ baryon, and $B$ the kaon or antikaon.

For the $AB$ interaction, we refer to chiral perturbation theory, in which the most important contribution is the leading order Weinberg-Tomozawa (WT) term~\cite{Altenbuchinger:2013vwa}
\begin{equation}
    V_{WT}(\mathbf{q}) = -\frac{C_W(I)}{2 f_\pi^2},
\end{equation}
where the pion decay constant $f_{\pi}=130$ MeV and $C_{W}(I)$ represents the strength of the WT interaction. This potential can be rewritten in coordinate space by Fourier transformation and we use the same form of the interaction as that adopted in Refs.~\cite{Wu:2019vsy,Wu:2020job,Wu:2020rdg}, which  explicitly reads
\begin{equation}
\label{Poten:DK}
    V_{AB}(r;R_c)=C(R_c)e^{-(r/R_c)^2}.
\end{equation}
Here $R_C$ is a coordinate space cutoff representing the effective interaction range. In this work, we choose $R_c$ ranging from 0.5 to 2.0 fm to study the related uncertainties. The $C(R_c)$ is a running coupling constant related to $R_c$. We determine the $C(R_c)$ of the $DK$ interaction by reproducing the $D_{s0}^*(2317)$ state. 
According to heavy quark symmetry~\cite{Neubert:1993mb}, the $B\bar{K}$ and $\Xi_{cc}\bar{K}$ interactions are the same as  the $DK$ one.

For the interactions between the two identical hadrons, we resort to phenomenological models, e.g., the one boson exchange (OBE) model developed in Ref.~\cite{Liu:2019stu}.
In Ref.~\cite{Wu:2019vsy}, the $DD$ OBE potential has been derived with the exchange of $\sigma$, $\rho$ and $\omega$ mesons. According to heavy quark flavor symmetry, the $BB$ interaction is the same as the $DD$ one. For the explicit form of the $DD$ OBE potential, see Ref.~\cite{Wu:2019vsy}. For the $\Xi_{cc}\Xi_{cc}$ potential in the OBE model, we can derive from the $DD$ OBE potential and the heavy antiquark-diquark symmetry (HADS)~\cite{Wu:2020rdg}.

\section{Gaussian expansion method}
\label{sec:GEM}
 
As all the two-body interactions have been specified, we use the GEM to solve the Schr\"odinger equation. In this section, we explain how to use the Gaussian expansion method (GEM) to study the  $DDK$, $BB\bar{K}$ and $\Xi_{cc}\Xi_{cc}\bar{K}$ systems.

 \begin{figure}[!h]
 	\centering
 	\includegraphics[width=8cm]{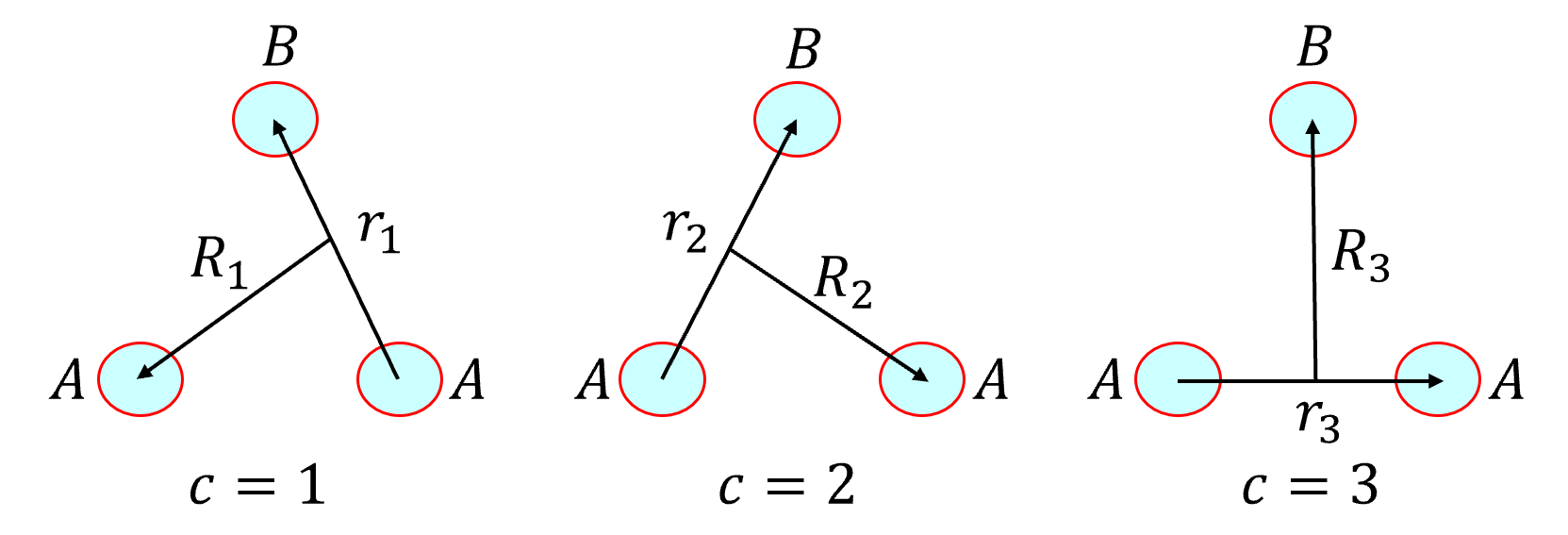}
 	\caption{Jacobi coordinate permutations for a three-body $AAB$ system.}
 	\label{Jac3body}
 \end{figure}

We study the three-body $AAB$ systems by solving the following Schr\"{o}dinger equation with three Jacobi coordinates shown in Fig.~\ref{Jac3body}
\begin{equation}\label{schd}
  \hat{H}\Psi_{JM}^{total}=E\Psi_{JM}^{total},
\end{equation}
with the Hamiltonian
\begin{equation}\label{hami}
  \hat{H} = \sum_{i=1}^{3} \frac{p_i^2}{2m_i}-T_{c.m.} +
   V_{AB}(r_1) + V_{AB}(r_2) + V_{AA}(r_3),
\end{equation}
where $T_{c.m.}$ is the kinetic energy of the center of mass and $V(r)$ is
the potential between the two relevant particles.
The three-body wave functions can be constructed in Jacobi coordinates as 
\begin{equation}
    \Psi_{JM}^{total}=\sum_{c=1}^{3}\Psi(\bf{r}_c,\bf{R}_c),
\end{equation}
where $c=1-3$ is the label of the Jacobi channels shown in Fig.~\ref{Jac3body}. In each Jacobi channel the wave function $\Psi(\mathbf{r}_c,\mathbf{R}_c)$ reads
\begin{equation}
    \Psi(r_c,R_c)=C_{c,\alpha}H^c_{T,t}\Phi_{lL,\lambda}(\mathbf{r}_c,\mathbf{R}_c)
\end{equation}
where $C_{c,\alpha}$ is the expansion coefficient and the $\alpha=\{nN,tT,lL\lambda\}$ labels the basis number with the configuration sets of the Jacobi channels. Here $n(N)$  is the number of Gaussian basis used and $l(L)$ is the orbital angular momentum corresponding to the Jacobi coordinates $r (R)$, $\lambda$ is the total orbital angular momentum coupled from $l$ and $L$. $H^c_{T,t}$ is the  three-body isospin wave function where $t$ is the isospin of the subsystem in Jacobi channel $c$ and $T$ is the total isospin.

The three-body spatial wave function $\Phi(\mathbf{r}_c,\mathbf{R}_c)$ is constructed by two two-body wave functions as
\begin{equation}
\begin{split}
       \Phi_{lL,\lambda}(\mathbf{r}_c,\mathbf{R}_c)&=[\phi_{n_cl_c}^{G}(\mathbf{r}_c)\psi_{N_cL_c}^{G}(\mathbf{R}_c)]_{\lambda},\\
     \phi_{nlm}^{G}(\mathbf{r}_c)&=N_{nl}r_c^le^{-\nu_n r_c^2} Y_{lm}({\hat{r}}_c),\\
     \psi_{NLM}^{G}(\mathbf{R}_c)&=N_{NL}R_c^Le^{-\lambda_n R_c^2} Y_{LM}({\hat{R}}_c).
\end{split}
\end{equation}
Here $N_{nl}(N_{NL})$ is the normalization constant of the Gaussian basis.

Considering that there are two identical particles in these 3-body systems, the total wave function should be symmetric(or antisymmetric) with respect to the exchange of the two identical particles, which requires
\begin{equation}\label{exchange}
  \hat{P}_{12}  \Psi_{JM}^{total}= P\Psi_{JM}^{total},
\end{equation}
where $\hat{P}_{12}$ is the exchange operator of particles 1 and 2, $P_{12}=+1$ for mesons and $P_{12}=-1$ for baryons.
 Considering only $S$-wave interactions and the constraint of the two identical particles, the quantum numbers of the $AAB$ systems are $I(J^P)=\frac{1}{2}(0^-)$.

\section{Results and discussions}

With the interaction inputs presented in Sec.~\ref{Interactions}, we study the  three $AAB$ systems, i.e., the $DDK$, $BB\bar{K}$, and $\Xi_{cc} \Xi_{cc} \bar{K}$ systems with the Gaussian expansion method.

\begin{table}[!h]
  \centering
    \caption{Binding energies (in units of MeV) and root mean square radii (in units of fm) of the $DDK$ bound state. }\label{BE:3body-1}
  \begin{tabular}{c c c c}
     \hline
     \hline
$R_c$ &$B_3(DDK)$&$ r_{DD}$&$ r_{DK}$\\
\hline
0.5&74.6 &1.08 &1.02  \\
1& 71.2&1.36 &1.32  \\
2& 68.8, 45.1&1.80 &1.80  \\
    \hline
    \hline
   \end{tabular}
\end{table}

Our results show that these  three-body systems are indeed bound. In Table~\ref{BE:3body-1}. We present the binding energies and root mean square (RMS) radii of the $DDK$ bound state. The binding energy of the $DDK$ bound state ranges from 68.8 to 74.6 MeV with the cutoff ranging from 0.5 to 2 fm, from which one can see that it is only weakly cutoff dependent. The RMS radius of $DK$ in the $DDK$ bound state ranges from 1.02 to 1.80 fm and that of $DD$ ranges from 1.08 to 1.80 fm as the cutoff increases. The RMS radii are strongly cutoff dependent because the cutoff $R_c$ represents the effective interaction range.

\begin{table}[!h]
  \centering
    \caption{Binding energies (in units of MeV) and root mean square radii (in units of fm) of the $BB\bar{K}$ bound states. }\label{BE:3body-2}
  \begin{tabular}{c c c c }
     \hline
     \hline
$R_c$ &$B_3(BB\bar{K})$&$ r_{BB}$&$ r_{B\bar{K}}$\\
\hline
0.5& 152.3, 74.2&0.71 &0.58 \\
1&108.5, 64.2&1.05 &0.83  \\
2& 87.3, 63.8&1.53 & 1.15 \\
    \hline
    \hline
   \end{tabular}
\end{table}
From heavy quark flavor symmetry, we know that the $B\bar{K}$ interaction is the same as the $DK$ one, as a result, the $BB\bar{K}$ system is analogous to the $DDK$ system.
Compared with the $DDK$ system, the heavier mass if the bottom quark causes two differences for the $BB\bar{K}$ system, see Table~\ref{BE:3body-2}. First, there are two bound states in the $BB\bar{K}$ system. There is also an exited state in the $DDK$ system with $R_c=2$ fm, but this state vanishes as the cutoff becomes smaller. Second, the binding energy of the $BB\bar{K}$ ground state is larger and strongly dependent on the cutoff, which is about 87 to 152 MeV as the cutoff decreases. Correspondingly, the RMS radius of the $B\bar{K}$ pair is about 0.58 to 1.15 fm and that of the $BB$ pair is about 0.71 to 1.53 fm, in the $BB\bar{K}$ ground state.

\begin{table}[!h]
  \centering
    \caption{Binding energies (in units of MeV) and root mean square radii (in units of fm) of the $\Xi_{cc}\Xi_{cc}\bar{K}$ bound state. }\label{BE:3body-3}
  \begin{tabular}{c c c c }
     \hline
     \hline
$R_c$ &$B_3(\Xi_{cc}\Xi_{cc}\bar{K})$&$ r_{\Xi_{cc}\Xi_{cc}}$&$ r_{\Xi_{cc}\bar{K}}$\\
\hline
0.5&118.4 &0.75 &0.80 \\
1&92.8 &1.04 &1.14  \\
2&79.7, 55.5 &1.43 &1.63 \\
    \hline
    \hline
   \end{tabular}
\end{table}

In Table~\ref{BE:3body-3}, we present the binding energies and RMS radii of the $\Xi_{cc} \Xi_{cc} \bar{K}$ bound state we predicted.
The $\Xi_{cc}$ baryon is the heavy anti-quark diquark symmetry (HADS) partner of the $D$ meson, thus the $\Xi_{cc} \Xi_{cc} \bar{K}$ bound state could be viewed as the HADS counterpart of the $DDK$ bound state.
The binding energy of the $\Xi_{cc} \Xi_{cc} \bar{K}$ bound state ranges from 79.7 to 118.4 MeV, of which the RMS radii of $\Xi_{cc}\bar{K}$ and $\Xi_{cc} \Xi_{cc}$ pairs are 0.80 to 1.63 fm and 0.75 to 1.43 fm, respectively.


There is  a remarkable phenomenon that can occur in a 3-body system, the Efimov effect, discovered by Efimov in 1970~\cite{Efimov:1970zz}. The Efimov effect refers to the appearance of a geometric spectrum in the 3-body system when at least two of the three pairs of particles are in the unitary limit, i.e. their scattering lengths diverge. In this work, the three $AAB$ systems we studied above do not have divergent scattering lengths, but the interactions between the subsystems are cutoff dependent, where the cutoff represents the effective interaction range. This is complementary to the Thomas collapse~\cite{Thomas:1935zz}: reducing the range of the interaction is equivalent to a relative increase of the scattering length when expressed in units of the range. Actually we can use the Efimov effect as a proxy to show the existence of the Thomas collapse in these systems, as proposed in other works~\cite{Valderrama:2018sap,Valderrama:2018azi,Wu:2020rdg}.

In the $AAB$ system, where $A$ and $B$ are two different species of
particles and the $AB$ interaction is resonant,
the condition for having the Efimov effect is
\begin{eqnarray}
  \lambda_{\alpha} = \frac{\sin{2 \alpha}}{2 \alpha} \leq \lambda \, ,
\end{eqnarray}
with $\lambda=1/2$ a geometric factor depending on the characteristics of
the $AB$ interaction and quantum numbers of the system. 

For the three $AAB$ systems we studied, $\lambda_{\alpha}$ of the $\Xi_{cc} \Xi_{cc} \bar{K}$ and $BB\bar{K}$ systems are 0.389 and 0.321 respectively~(for details, see Ref.~\cite{Wu:2020rdg}), from which we have  $\lambda \geq \lambda_{\alpha}$: the conclusion is that for the $\Xi_{cc} \Xi_{cc} \bar{K}$  and $BB\bar{K}$ systems the Effimov
effect can indeed happen. But of course, from the fact that the $AB$ system is far from the divergence of the scattering length, what we can  expect is Thomas collapse~\cite{Thomas:1935zz}.
\begin{figure}[!h]
 	\centering
 	\includegraphics[width=8cm]{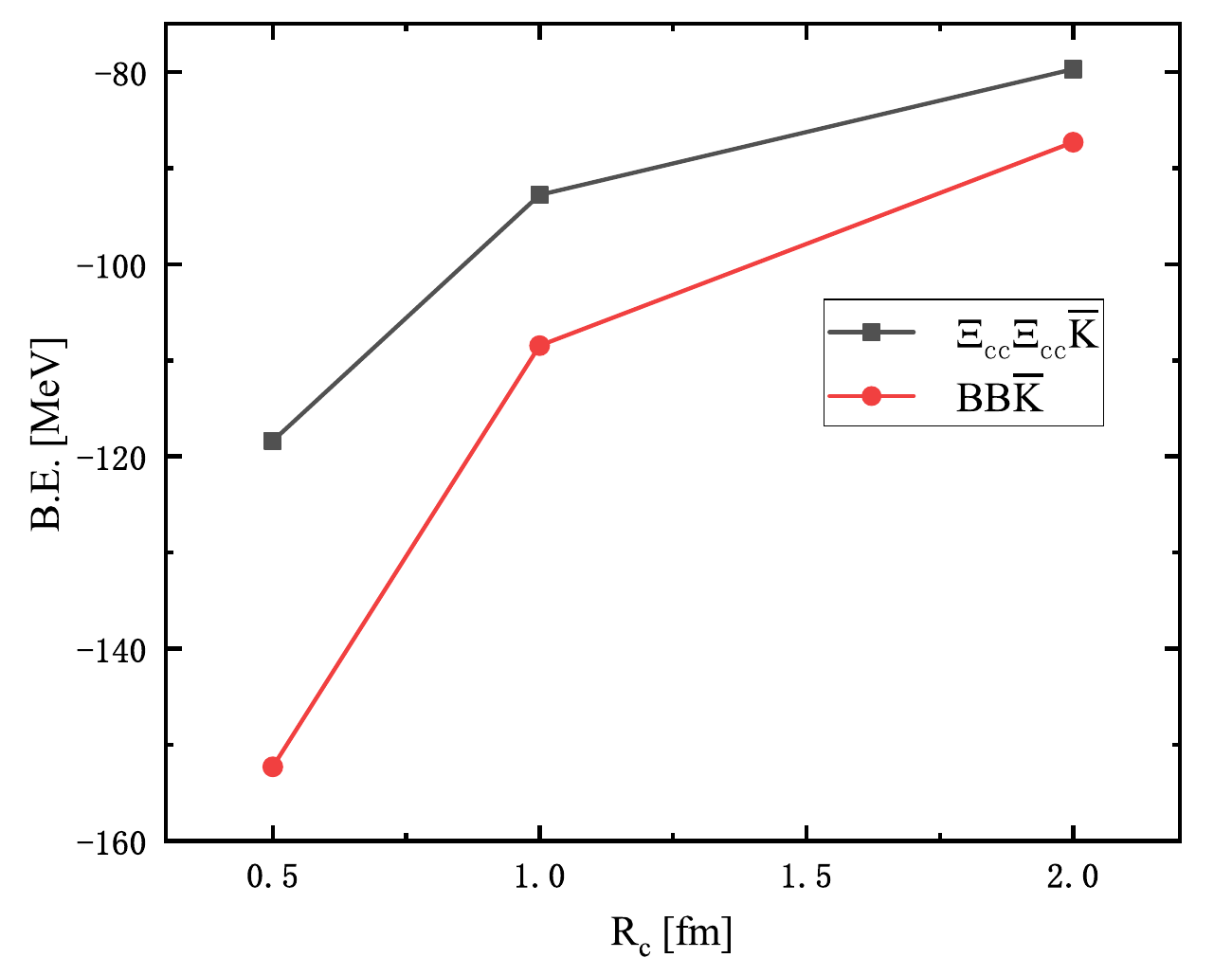}
 	\caption{Thomas collapse of the $\Xi_{cc} \Xi_{cc} \bar{K}$ and $BB\bar{K}$ systems.}
 	\label{TC}
 \end{figure}
This is what we indeed obtain. Our results show that the binding energies of the
$\Xi_{cc} \Xi_{cc} \bar{K}$ and $BB\bar{K}$ systems are strongly cutoff dependent. More specifically, the binding energies become divergent as the cutoff $R_c$ goes to 0, see Fig.~\ref{TC}. The results clearly show the Thomas collapse in the $\Xi_{cc} \Xi_{cc} \bar{K}$ and $BB\bar{K}$ systems.
As a comparison, $\lambda_{\alpha}$ of the $DDK$ system is 0.531, from which we deduce that there is no Thomas collapse in this case, which are consistent with our results shown in Table~\ref{BE:3body-1}.

\section{Summary}
Based on the molecular picture of two-body hadronic states and heavy quark symmetry, we studied three-body $AAB$ systems, i.e., $DDK$, $BB\bar{K}$, and $\Xi_{cc} \Xi_{cc} \bar{K}$, and found that they indeed bind.
In these $AAB$ systems, if the mass differences of the $A$ and $B$ particles are large enough, such as the $\Xi_{cc} \Xi_{cc} \bar{K}$ and $BB\bar{K}$ systems, there could exist Thomas collapse, which indicates that the three-body binding energy become divergent as the interaction range of $AB$ goes to zero.

These predicted bound states are expected to be observed in the current or updated experimental facilities. If they are found, the picture of two-body hadronic molecules can be tested and supported in a highly non-trivial way.



\bibliographystyle{spphys} 
\bibliography{mybibs}
\end{document}